\documentclass[conference,10pt]{IEEEtran}

\usepackage[subrefformat=parens,labelformat=parens,caption=false,font=footnotesize]{subfig}
\usepackage{array}
\usepackage{amsmath}
\usepackage{breqn}
\usepackage[nolist,nohyperlinks]{acronym}
\usepackage{graphicx,wrapfig,lipsum}
\usepackage{rotating}
\usepackage{cleveref}
\usepackage{epstopdf}
\usepackage{balance}
\usepackage{amsmath, amssymb, amsthm}
\usepackage{stmaryrd}
\usepackage{verbatim}
\usepackage[short]{optidef}
\usepackage{url}
\usepackage[utf8]{inputenc}
\usepackage[english]{babel}

\usepackage{multicol}
\usepackage{xr-hyper}

\usepackage{scalerel}
\usepackage{multicol}

\usepackage[noadjust]{cite}
\usepackage[normalem]{ulem}
\usepackage{color}
\usepackage{dsfont}
\usepackage{bm}
\usepackage[short]{optidef}
\usepackage{diagbox}
\usepackage{enumitem}
\usepackage{slashbox}
\usepackage[ruled]{algorithm2e}
\usepackage{multirow}
\usepackage[T1]{}
\usepackage{color, colortbl}
\usepackage{babel}
\usepackage{verbatim}
\usepackage{textcomp}
\usepackage{tabulary}
\usepackage{caption}
\usepackage{bbm}
\definecolor{Gray}{gray}{0.95}
\definecolor{LightCyan}{rgb}{0.8,0.85,1}
\definecolor{LightBlue}{rgb}{0.6,0.6,1}
\usepackage[font=small]{caption}

\setlist{nosep}
\newcommand\blfootnote[1]{%
  \begingroup
  \renewcommand\thefootnote{}\footnote{#1}%
  \addtocounter{footnote}{-1}%
  \endgroup
}
\usepackage{mathtools}

\begin{document}
\title{On the Role of Non-Terrestrial Networks for Boosting Terrestrial Network Performance in Dynamic Traffic Scenarios}

\author{Henri Alam$^{\dag\ddag}$, Antonio De Domenico$^\dag$, Florian Kaltenberger$^\ddag$, David L\'{o}pez-P\'{e}rez$^\bigstar$  \\
 \small{$^\dag$Huawei Technologies, Paris Research Center, 20 quai du Point du Jour, Boulogne Billancourt, France.} \\
 \small{$^\ddag$EURECOM, 2229 route des Crêtes, 06904 Sophia Antipolis Cedex, France.} \\
 \small{$^\bigstar$Universitat Politècnica de València, Spain.}
 }
\maketitle

\thispagestyle{empty}

\begin{abstract}
Due to an ever-expansive network deployment, numerous questions are being raised regarding the energy consumption of the mobile network.
Recently, Non-Terrestrial Networks (NTNs) have proven to be a useful, and complementary solution to Terrestrial Networks (TN) to provide ubiquitous coverage.
In this paper, we consider an integrated TN-NTN, and study how to maximize its resource usage in a dynamic traffic scenario.
We introduce BLASTER, a framework designed to control User Equipment (UE) association, Base Station (BS) transmit power and activation, and bandwidth allocation between the terrestrial and non-terrestrial tiers. 
Our proposal is able to adapt to fluctuating daily traffic, focusing on reducing power consumption throughout the network during low traffic and distributing the load otherwise.
Simulation results show an average daily decrease of total power consumption by $45 \%$ compared to a network model following 3GPP recommendation, as well as an average throughput increase of roughly $250 \%$.
Our paper underlines the central and dynamic role that the NTN plays in improving key areas of concern for network flexibility. 
\blfootnote{This research is supported by the Generalitat Valenciana through the CIDEGENT PlaGenT, Grant CIDEXG/2022/17, Project iTENTE, and by the action CNS2023-144333, financed by MCIN/AEI/10.13039/501100011033 and the European Union “NextGenerationEU”/PRTR.}
\end{abstract}

\section{Introduction}
\label{sec:intro}

With the swift evolution of cellular communications in recent years, there has been a significant upswing in the demand for high-speed data connectivity.
This has led to the imposition of strict prerequisites to deliver high capacity and ensure ubiquitous connectivity for the network.
The use of heterogeneous networks (HetNets) has been proposed as a solution to address these requirements \cite{Xu_2021}. Indeed, the use of HetNets creates a multi-layered network, allowing efficient data offloading which improves both the capacity and the coverage throughout the network.
However, this dense deployment increases the overall energy consumption of the network, which is not desirable given the current environmental and economic context.
Hence, one of the key objectives of deploying and operating mobile networks is to reduce power usage while ensuring to meet Quality of Service (QoS) requirements \cite{Lopez_2022}. 
In the past few years, non-terrestrial networks (NTNs) have arisen as a feasible approach to supplement the terrestrial network (TN) and extend coverage to previously underserved geographic regions \cite{Ahmmed_2022}. 
An NTN is a network in which airborne vehicles like drones (i.e. unmanned aerial vehicles), high-altitude platform stations (HAPS), or satellites function as a relay node or base station (BS) to provide connectivity for each user equipment (UE) within the network.
The inherent advantage of NTNs lies in their ability to offer extensive coverage across vast regions, including remote geographical areas where deploying terrestrial macro BSs (MBSs) would be cost-prohibitive or logistically challenging. Among the various deployment possibilities, it appears that low-earth orbit (LEO) satellites will lead the way in achieving high-capacity connectivity from space \cite{Giordani_2021}. They orbit at altitudes ranging from 200 to 2000 kilometers.
Due to their closer proximity to Earth, they offer superior signal strength and reduced latency when compared to alternative satellite designs. This results in lower energy requirements for launch and decreased power consumption for signal transmission from/to the satellite.
Considering all of these factors, integrated TN-NTN may be the path towards efficient service of both terrestrial and aerial UEs \cite{Benzaghta_2022}.
Typically, each UE is connected to the BS with the strongest recorded Reference Signal Received Power (RSRP) within the network. However, this association strategy has its limits as it does not consider the varying traffic demands of UEs, potentially resulting in suboptimal load balancing and subsequent performance issues. An improved UE association policy should not only consider the strength and quality of the UE's signal but also factor in the load on each cell.
Our previous work in \cite{Alam_2023_2} addresses this specific problem in an integrated TN-NTN. We were able to distribute the network load with the help of satellites and enhance the maximum network throughput as well as improve network coverage.\newline 
Regarding Energy Efficiency (EE), we acknowledge that utilizing all MBSs during low-traffic scenarios (e.g. night) might not be ideal. Indeed, most of them may be under-utilized or not used at all, resulting in an inefficient allocation of energy and communication resources.
Hence, given the context of an integrated TN-NTN, it may be appropriate to turn off part of the terrestrial MBSs and offload the UEs to the satellites to reduce energy consumption.
To the extent of our knowledge, most of the work related to BS activation does not consider the NTNs as a solution to the problem of maintaining coverage and capacity requirements while shutting down MBSs.
The authors of \cite{Oh2013} implement a switching-on/off-based energy-saving algorithm that shuts down the MBSs one by one while making sure that they do not overload the neighbouring MBSs.
To maintain the user experience, some of the works have considered minimal QoS requirements. In \cite{Chen2015}, the authors study the impact of traffic offloading in HetNets on energy consumption and suggest a centralized Q-learning approach for achieving a balance between conserving energy and ensuring QoS satisfaction. 
The authors of \cite{Shen_2017_Letter} develop an algorithm that allows each UE to associate with multiple MBSs on different frequency bands, and optimize the MBS's transmit power simultaneously so that they can be shut down during low traffic. 
\newline In this paper, we present a Bandwidth spLit, user ASsociaTion and PowEr contRol framework (BLASTER). It is a novel adaptive radio resource management framework, which controls the bandwidth split, UE association, terrestrial MBS transmit power and activation in an integrated TN-NTN.
It is able to trade-off between network capacity and network energy consumption based on the traffic state.
Our results show that the proposed framework reduces the average network power consumption by $45 \%$ compared to an integrated TN-NTN model following 3GPP recommendations, while significantly improving the mean throughput during high-traffic periods.



\section{System Model}
We study a downlink (DL) cellular network comprising $M$ terrestrial MBSs and $N$ MBSs mounted on LEO satellites, totalling $L$, all of which can serve $K$ UEs deployed in a rural area. We denote the total system bandwidth as $W$, which the mobile network operator allocates between the terrestrial and non-terrestrial tiers. In our research, we assume that this network operates in the S band at approximately 2 GHz, with terrestrial and satellite MBSs utilizing orthogonal portions of it. Throughout the remainder of this paper, we will use $\mathcal{T}$ (representing terrestrial) and $\mathcal{S}$ (representing satellites) to denote the sets of MBSs. Furthermore, let $\mathcal{U} =  \{1, \cdots, i, \cdots,K \}$ be the set of UEs and $\mathcal{B} = \mathcal{T} \cup \mathcal{S} = \{ 1, \cdots, j, \cdots, L \} $ the set of all MBSs. Concerning the channel model, the large-scale channel gain between a terrestrial MBS $j$ and a UE $i$ is computed as follows:
\begin{equation}\label{channel_terrestrial}
     \beta_{ij}  = G_{T_X} \cdot PL_{ij} \cdot SF_{ij} ,
\end{equation}
where $G_{T_X}$ is the transmit antenna gain, 
$PL_{ij}$ is the path loss, 
and $SF_{ij}$ is the shadow fading. Note that the UE antenna gain is $0$ dBi.
Conversely, when a satellite MBS $j$ serves a UE $i$, the large-scale channel gain is the following \cite{3GPPTR38.811}:
\begin{equation}\label{channel_satellite}
     \beta_{ij}  = G_{T_X} \cdot PL_{ij} \cdot SF_{ij} \cdot CL \cdot PL_s.
\end{equation}
In \eqref{channel_satellite}, $CL$ represents clutter loss, an attenuation due to buildings and vegetation near the UE, while $PL_s$ accounts for scintillation loss, which encompasses rapid fluctuations in signal amplitude and phase caused by ionospheric conditions.
Given that each UE is exclusively served by either a terrestrial or a satellite MBS, and both tiers operate without interference due to their orthogonal bandwidth allocation, we can compute the large-scale Signal-to-Interference-plus-Noise Ratio (SINR) for each UE $i$ as outlined below:
\begin{equation}\label{SINR}
    \gamma_{ij}  = \frac{ \beta_{ij} p_j}{ \sum\limits_{\substack{ j^' \in \mathcal{I}_j}} \beta_{ij^'} p_{j^'} + \sigma^2},
\end{equation}
where $p_j$ is the transmit power allocated per resource element (RE) at MBS $j$, 
$\mathcal{I}_j$  represents the set of MBSs causing interference to the serving MBS $j$,
and $\sigma^2$ is the noise power per RE.
Subsequently, if we assume that MBS $j$ evenly allocates its available bandwidth $W_j$ among the $k_j$ UEs it serves, we can calculate the average throughput for UE $i$ connected to MBS $j$ as follows:
\begin{equation}\label{Shannon_data_rate}
    R_{ij}  = \frac{W_j}{k_j} \log_2(1 + \gamma_{ij}).
\end{equation}
The power consumption model for a terrestrial MBS depends on multiple parameters, such as bandwidth, number of antennas, or power amplifier efficiency, as detailed in \cite{Piovesan2022}.
This model can be compacted as:
\begin{equation}\label{Power_conso_model_v1}
    Q_j(p_j) = P_0 + p_j + \psi_j \vert\vert p_j \vert \vert_0,
\end{equation}
\noindent where $\psi_j$ and $p_j$
respectively include all the static (transmit power independent) and dynamic (transmit power dependent) components of the model, and $\vert\vert \cdot \vert \vert_0$ is a binary-valued function equal to $1$ if the transmit power $p_j$ is greater than 0.
$P_{0}$ is the power consumption of the components that stay active in a shutdown terrestrial MBS.
We suppose that the power consumed by a satellite is harvested from solar panels.
In the following, please note that we will be referring to the Hadamard product using the symbol $\odot$.
 \section{Problem Formulation}
We aim to design a mechanism that jointly optimizes network capacity and TN energy consumption by dynamically adapting the resource allocation to the network load.
By maximizing the sum of the log-throughput (SLT) perceived by each UE, we want to ensure proportional fairness as the nature of the logarithm prevents excessive use of resources for a particular UE, promoting a proportional allocation of resources.
We define $\varepsilon$ as the portion of the bandwidth allocated for the LEO satellites.
Consequently, we can calculate the bandwidth $W_j$ for MBS $j$ as $W\varepsilon$ when it is a satellite, or as $W(1-\varepsilon)$ when it is a terrestrial MBS.
Let us also introduce a binary variable, denoted as $x_{ij}$, which takes the value of $1$ when UE $i$ is connected to MBS $j$ and $0$ otherwise.
We can then write the perceived throughput for UE $i$ as:
\begin{equation}\label{Shannon_data_rate_UE}
    R_{i}  =  \sum_{j \in \mathcal{B}} x_{ij}R_{ij.}
\end{equation}
Our target is to optimize the UE-BS association, the transmit power at each MBS, and the allocation of bandwidth to each tier to find the right trade-off between maximizing the network SLT and minimizing the terrestrial network power consumption, by shutting down as many terrestrial MBSs as possible.
This can be written as follows:
\begin{maxi!}|s|[2]
{X,\varepsilon, p}{\sum\limits_{i \in \mathcal{U}} \log(R_i) - \lambda \sum_{j\in\mathcal{B}} Q_j(p_j)}{}{}\label{OPT_PB_1}
\addConstraint{x_{ij}}{\in \{0,1\}, \; i \in \mathcal{U}, j \in \mathcal{B}}{}\label{PB1_const1}
\addConstraint{\tilde{\beta}\cdot p }{\geq RSRP_{\rm min} \cdot \mathbbm{1}_{K}, \; \forall i \in \mathcal{U}}{
}\label{PB1_const2}
\addConstraint{p_j}{\leq p_{j}^{MAX}, \; \forall j \in \mathcal{B}}\label{PB1_const3}
\addConstraint{\varepsilon}{ \in \left[ 0,1 \right], }{
}\label{PB1_const4}
 \end{maxi!}
where X $= \left[x_{ij}\right]$ is the binary association matrix, p $ = \left[ p_1, \dots, p_{L} \right]^T$ is the vector representing the transmit power at each MBS and $\tilde{\beta} = X \odot \beta$.
Also, $\lambda$ is a scaling parameter used to manage the trade-off between SLT and power consumption, and fixed prior to the optimization, accordingly to the expected user traffic.
Constraint \eqref{PB1_const2} guarantees that the minimum RSRP for each UE must exceed a predefined threshold $RSRP_{\rm min}$.
Additionally, constraint \eqref{PB1_const3} limits the maximum transmit power allocated per RE in each MBS $j$ to $p_j^{\rm MAX}$.
Since one of our aims is to reduce the terrestrial network power consumption, a sparse solution for the power vector $p$ would be ideal.
However, since the power consumption model \eqref{Power_conso_model_v1} contains a non-continuous term, it may be hard to optimize.
With this in mind, we approximate the utility function \eqref{OPT_PB_1} by introducing a mixed $L_1$-$L_2$ penalty function which promotes group sparsity, as done in \cite{Shi2013}.
We then have the following:
\begin{maxi!}|s|[2]
{X,\varepsilon, p}{\sum\limits_{i \in \mathcal{U}} \log(R_i) - \lambda \left( \vert\vert p\vert\vert_1 + \sum_{j=1}^{L} \psi_j w_j \vert\vert p\vert\vert_2 \right)}{}{}\label{OPT_PB_2}
\addConstraint{\eqref{PB1_const1} - \eqref{PB1_const4},}{}{}\label{PB2_const1}
\end{maxi!}
where $\vert\vert\cdot\vert\vert_1$, $\vert\vert\cdot\vert\vert_2$ represent the $L_1$ and $L_2$ norm and $w_j$ represents the power weight of MBS $j$. 
Those weights are inversely proportional to the transmit power of each MBS, hence pushing those with low transmit power to be shut down.
\section{Full Breakdown of BLASTER}
\label{sec:UtilityOptimization}

In this section, 
we study BLASTER, the framework proposed to solve the optimization problem \eqref{OPT_PB_2}-\eqref{PB2_const1}.
We use the block coordinate gradient ascent (BCGA) algorithm, meaning that we first optimize the UE-BS association and bandwidth allocation considering fixed transmit power, 
similarly to \cite{Alam_2023_2}. 
Thereafter, we optimize the transmit power level considering the first two parameters fixed. 

\subsection{Utility optimization under fixed transmit power}
Denoting by $f$ the utility function that we want to maximize in \eqref{OPT_PB_2}, we notice that we have a convex optimization problem with respect to $X$ ($\varepsilon$ and $p$ being fixed). We can then use the iterative gradient projection method to solve the problem, as it is particularly well-suited for constrained optimization problems.
It consists of computing the gradient and projecting it onto the feasible region defined by the constraints. 
Embracing the gradient projection method, we compute the gradient update at time-step $s$ as:
\begin{equation}\label{X_tilde}
    \tilde{X}(s) = X(s) + \alpha\nabla_{X}f\left(X,p,\varepsilon \right),
\end{equation}

\noindent where $\alpha \in \mathbb{R}^{K \times L }$ is a step-size chosen appropriately and $\nabla$ is the gradient operator.
Then, we write the projection step in the following form:
\begin{mini!}|s|[2]
{X(s)}{\frac{1}{2} \vert \vert X(s) - \tilde{X}(s) \vert \vert^2_F}{}{}\label{Projection_problem_X}
\addConstraint{\tilde{\beta} \cdot p \geq RSRP_{\rm min}\cdot \mathbbm{1}_{K},}{}{}\label{Projection_PB_Const1}
\end{mini!}
\noindent where $\vert\vert \cdot \vert \vert_F$ represents the Frobenius norm.
For the following, we simplify the notation by omitting time-step indices.
To solve \eqref{Projection_problem_X}-\eqref{Projection_PB_Const1}, we use the Lagrange multipliers method.
Therefore, we compute the Lagrangian function associated with the problem:
\begin{equation}\label{Lagrange_formula_projection}
\begin{split}
    \mathcal{L}\left( X,\mu \right) &= \frac{1}{2}  \vert \vert X - \tilde{X} \vert \vert^2_F + \left(\tilde{\beta} \cdot p - RSRP_{\rm min} \cdot \mathbbm{1}_{K}\right)^T \mu\\
    &= \frac{1}{2}  \vert \vert X \vert \vert^2_F - \mathrm{Tr}\left( X^T \tilde{X} \right) + \frac{1}{2}  \vert \vert \tilde{X} \vert \vert^2_F + \left( \tilde{\beta} \cdot p \right)^T \mu \\ 
    &- \left( RSRP_{\mathrm{min}} \cdot \mathbbm{1}_{K}\right)^T \mu,
\end{split}
\end{equation}
where $\mu \in \mathbb{R}^K$ is the Lagrange multiplier associated with constraint \eqref{PB1_const2}.
Computing the gradient of \eqref{Lagrange_formula_projection} with respect to $X$, we get:
\begin{equation}\label{Gradient_X_Lagrange_formula_projection}
\begin{split}
    \nabla_X \mathcal{L}\left(X,\mu \right) = X - \tilde{X} + \beta \odot \underbrace{ \left( \mathbbm{1}_{K} \cdot p^T \right)}_{:= p^{\rm PAD}} \odot \underbrace{\left( \mu \cdot \mathbbm{1}_{L}^T \right)}_{:=\mu^{\rm PAD}}.
\end{split}
\end{equation}
Then, we know that the optimal value to minimize \eqref{Lagrange_formula_projection} with a fixed dual variable is:
\begin{equation}\label{X_opt_formula}
\begin{split}
X^\star = \max \{ \tilde{X} - \beta \odot  p^{\rm PAD} \odot \mu^{\rm PAD}, 0 \}.
\end{split}
\end{equation}

\noindent Having computed $X^\star$, we thereby introduce the Lagrangian dual function, 
which can be written as:

\begin{equation}\label{Dual_definition}
\begin{split}
    &\mathcal{D}\left(\mu\right) = \max\limits_{X} \mathcal{L}\left(X,\mu \right).
\end{split}
\end{equation}
Hence, after injecting \eqref{X_opt_formula} into the formula above, we get:
\begin{equation}\label{Dual_definition_2}
\begin{split}
    \mathcal{D}\left(\mu\right) &= \mathcal{L}\left(X^\star,\mu \right) \\
    &= \frac{1}{2}  \vert \vert X^\star \vert \vert^2_F - \mathrm{Tr}\left( {X^\star}^T \tilde{X} \right) + \frac{1}{2}  \vert \vert \tilde{X} \vert \vert^2_F \\
    &+ \Big[ \big( X^\star \odot \beta  \big) \cdot p \Big]^T \mu - \left( RSRP_{\mathrm{min}} \cdot \mathbbm{1}_{K}\right)^T \mu.
\end{split}
\end{equation}

\noindent  By noticing that $$\Big[ \big( X^\star \odot \beta  \big) \cdot p \Big]^T \mu = \mathrm{Tr}\left(X^\star \left( \beta \odot  p^{\rm PAD} \odot \mu^{\rm PAD} \right)^T \right),$$ we can rewrite \eqref{Dual_definition_2} as:
\begin{equation}\label{Dual_definition_3}
\begin{split}
    \mathcal{D}\left(\mu\right) &= \frac{1}{2}  \vert \vert X^\star \vert \vert^2_F - \mathrm{Tr}\left( X^\star \left[ \tilde{X} - \beta \odot  p^{\rm PAD} \odot \mu^{\rm PAD} \right]^T \right) \\ 
    &- \left( RSRP_{\mathrm{min}} \cdot \mathbbm{1}_{K}\right)^T \mu.
\end{split}
\end{equation}

\noindent Also, as the authors of \cite{Shen_2017_Letter} demonstrated, we know that $$ \frac{1}{2} \Big\vert \Big\vert \max \{A,0\} \Big\vert \Big\vert^2_F - \mathrm{Tr}\left( \max \{ A,0\} A^T\right)  =  - \frac{1}{2} \Big\vert \Big\vert \max \{A,0\} \Big\vert \Big\vert^2_F
$$ for a given matrix $A$.

\noindent Therefore, we can write the dual problem associated with \eqref{Projection_problem_X} as the following:
\begin{mini!}|s|[2]
{\mu}{ \frac{1}{2} \vert \vert X^\star\vert \vert^2_F \  + \ \left( RSRP_{\mathrm{min}} \cdot \mathbbm{1}_{K}\right)^T \mu }{}{}\label{Projection_dual_problem}
\addConstraint{\mu \leq 0}{}{}\label{Projection_dual_problem_const1}
\end{mini!}
Since this problem has a sole constraint, and we know that the projection onto the non-positive orthant is a straightforward task, we can use the gradient projection method to solve it.
After obtaining the solution to the problem above, we obtain $\mu^*$ and can retrieve the optimal solution to our projection problem \eqref{Projection_problem_X}-\eqref{Projection_PB_Const1}:
\begin{equation}\label{X_s+1}
X(s+1) \triangleq X^\star = \max \{ \tilde{X}(s) - \beta \odot  p^{\rm PAD} \odot {\mu^{*}}^{\rm PAD}, 0 \}.    
\end{equation}
After the convergence of this method, we are able to find the optimal association $X^*$. Afterwards, we need to split the bandwidth between both tiers in an optimal way. To drive the process, we introduce $r_{ij}$ as:
$$r_{ij} = \frac{W}{k_j} \log_2\left( 1 + \gamma_{ij} \right)$$ such that:
\begin{equation}
\begin{split}
    R_{ij} =
    \begin{cases}
      \varepsilon r_{ij} & \text{if} \ j\in \mathcal{S},\\
      \left(1 - \varepsilon \right) r_{ij} & \text{otherwise}.
    \end{cases}
\end{split}
\end{equation}
\newline 
As both the transmit power allocated by the cells and the noise power linearly increase with the bandwidth, $\gamma_{ij}$ does not depend on it.
Therefore, we can compute the gradient of our utility function $f$ with respect to $\varepsilon$:
\begin{equation}\label{Gradient_f_epsilon}
\begin{split}
      & \nabla_\varepsilon f\left(X,p,\varepsilon \right) = \frac{\partial}{\partial \varepsilon} \left( \sum_{i =1}^K \log \left( R_i \right)  \right) 
      = \sum_{i =1}^K \frac{\partial}{\partial \varepsilon} \log \left( R_i \right) \\
      &= \sum_{i =1}^K \frac{ \sum\limits_{j \in \mathcal{S}}  \frac{\partial}{\partial\varepsilon} \left[ \varepsilon x_{ij}r_{ij} \right] + \sum\limits_{j \in \mathcal{T}}  \frac{\partial}{\partial\varepsilon} \left[ \left( 1 - \varepsilon \right) x_{ij}r_{ij} \right]   }{R_i}\\
      & \Leftrightarrow \nabla_\varepsilon f\left(X,p,\varepsilon \right) = \sum_{i =1}^K \left[ \frac{\sum\limits_{j \in \mathcal{S}} x_{ij}r_{ij} - \sum\limits_{j \in \mathcal{T}} x_{ij}r_{ij}}{R_i}   \right]
\end{split}
\end{equation}
By noticing that $\mathcal{U} = \mathcal{U_S}\ \cup\ \mathcal{U_T}$, where $\mathcal{U_S}$ and $\mathcal{U_T}$ represent the set of UEs served by the satellites and terrestrial MBSs respectively, we are able to find the optimal split by forcing \eqref{Gradient_f_epsilon} to $0$:
\begin{equation}\label{optimal_epsilon}
\begin{split}
        &\nabla_\varepsilon f\left(X,p,\varepsilon \right) = 0 \Leftrightarrow \sum_{i \in \mathcal{U_S}} \frac{1}{\varepsilon} + \sum_{i \in \mathcal{U_T}} \frac{-1}{1 - \varepsilon} = 0 \\ 
        &\Leftrightarrow \frac{K_{\mathcal{S}}}{\varepsilon} - \frac{K - K_{\mathcal{S}}}{1 - \varepsilon} = 0 \Leftrightarrow \varepsilon^* = \frac{K_{\mathcal{S}}}{K}
\end{split}
\end{equation}
\vspace{-0.5em}
where $K_{\mathcal{S}}$ represents the number of UEs associated to a satellite in the network.
\noindent From eq. \eqref{optimal_epsilon}, we notice that the share of allocated bandwidth for the non-terrestrial tier correlates intuitively with the number of UEs associated to a satellite.
\subsection{Transmit power optimization under fixed association}

After addressing the UE-BS association and bandwidth allocation challenges, we fix X and $\varepsilon$ to fine-tune the transmit power at each terrestrial MBS and ultimately, maximize the utility function. 
The transmit power optimization problem can then be expressed as:
\vspace{-0.5em}
\begin{maxi!}|s|[2]
{p}{\sum\limits_{i \in \mathcal{U}} \log(R_i) - \lambda \left( \vert\vert p\vert\vert_1 + \sum_{j=1}^{L} \psi_j w_j \vert\vert p\vert\vert_2 \right)}{}{}\label{OPT_PB_3}
\addConstraint{\eqref{PB1_const2} - \eqref{PB1_const3}}{}{}\label{PB3_const1}
\end{maxi!}
Due to the non-smoothness of the $L_1$ norm, we have to resort to the iterative proximal gradient method \cite{Boyd_2014} to solve \eqref{OPT_PB_3}-\eqref{PB3_const1}.
The gradient update at time-step $s$ is computed as such:
\begin{equation}\label{Gradient_power}
\tilde{p}(s) = p(s) + \eta \nabla_p f\left(X,p,\varepsilon \right)
\end{equation}
where $\eta \in \mathbb{R}^{L}$ is a step-size chosen appropriately.
As demonstrated in \cite{Boyd_2014}, the proximal gradient method updates $p$ by solving the following problem:
\begin{mini!}|s|[2]
{p}{\frac{1}{2} \vert \vert \tilde{p}(s) - p(s) \vert \vert^2_2 + t \vert \vert p(s) \vert \vert_2}{}{}\label{Proximal_problem}
\end{mini!}
where 
\begin{equation}\label{t}
    t = \lambda \cdot \eta \cdot w^T\psi.
\end{equation}
\noindent This problem has a closed-form solution, referred to as block soft thresholding \cite[Sec. 6.5.1]{Boyd_2014}:
\begin{equation}\label{Solution_proximal_gradient}
\hat p(s) = \rm\max \bigl\{ 1 - \frac{t}{\vert\vert \tilde{p}(s)
 \vert \vert_2}, 0 \bigr\}\tilde{p}(s).
\end{equation}
Once we have updated the transmit power vector, we must project it into a feasibility region which would ensure that constraints \eqref{PB1_const2} and \eqref{PB1_const3} are respected.\newline
Naturally, the upper bound of our feasible region is the maximum transmit power per RE.
To establish the lower limit, we use the minimal coverage constraint. In fact, from \eqref{PB1_const2} we know that each UE associated with a MBS $j$ should experience a signal power level exceeding the threshold value of $RSRP_{\rm min}$.
This can be rewritten as:
\begin{equation}
\begin{split}
& \forall i \in \mathcal{U}_j,\quad p_j \quad {\geq} \quad \frac{RSRP_{\rm min}}{ \beta_{ij}},
\end{split}
\end{equation}
with $\mathcal{U}_j$ being the set of UEs associated to the MBS $j$.
We are therefore able to establish the lower bound of the feasibility region for each MBS $j$ as:
\begin{equation}\label{tau_j}
\tau_j=\max _{i \in \mathcal{U}_j}\left(\frac{RSRP_{\min }}{\beta_{ij}}\right).
\end{equation}

\noindent Finally, the transmit power update done at the end of step $s$ is written as such:
\begin{equation}\label{power_update}
p{(s+1)}=\Bigr[ \hat{p}(s) \Bigr]_{\tau_j}^{p^{\rm MAX}}.
\end{equation}
\vspace{-0.2em}
Once the algorithm has converged and we obtain $p^*$, we update the power weights $w_j$ according to the re-weighting algorithm detailed in \cite{Shen_2017_Letter}.
The full optimization framework is summarized below in Algorithm \ref{Algorithm}.

\begin{algorithm}
\caption{BLASTER Framework}\label{Algorithm}
\KwData{K UEs and L MBs.}
Initialization\;
s = $0$\;
X: Association done through max-RSRP\;
p: Transmit power set to maximum\;
$\varepsilon = 0.5$\tcp*{Equal bandwidth split}
\textbf{Compute:} $f\left(X,\varepsilon,p\right)$ \tcp{Initial point}
$w = \left[1,\dots,1\right] \in \mathbb{R}^L$\;
Initialize $\alpha \in \mathbb{R}^{K\times L}$, $\mu \in \mathbb{R}^K$\;
Initialize $\eta \in \mathbb{R}^L$\;
\While{Utility function $f$ has not converged}{
    \tcp{UE Association and bandwidth split}
        
        \textbf{Compute:} $\tilde{X}(s) = X(s) + \alpha\nabla_{X}f\left(X,p,\varepsilon \right)$ \hspace*{0.18cm}(\ref{X_tilde})\;
        Solve (\ref{Projection_dual_problem}) using gradient projection to obtain $\mu^*$\;
        \textbf{Compute:} \\ $X(s+1) = \max \{ \tilde{X}(s) - \beta \odot  p^{\rm PAD} \odot {\mu^{*}}^{\rm PAD}, 0 \} (\ref{X_s+1})$\;
    $\varepsilon^* = \frac{K_{\mathcal{S}}}{K}$ \hspace*{5.2cm}(\ref{optimal_epsilon})\;
    \tcp{Power control step}
        \textbf{Compute:} $\tilde{p}(s) = p(s) + \eta \nabla_p f\left(X,p,\varepsilon \right) \quad $ \hspace*{\fill} (\ref{Gradient_power})\;
        \textbf{Compute:} $t = \lambda \cdot \eta \cdot w^T\psi \quad$ \hspace*{2.3cm}(\ref{t})\;
        \textbf{Compute:} $\hat p(s) = \rm\max \bigl\{ 1 - \frac{t}{\vert\vert \tilde{p}(s) \vert \vert_2}, 0 \bigr\}\tilde{p}(s)$\hspace*{0.2cm} (\ref{Solution_proximal_gradient})\;
        \textbf{Compute:} $\tau$ based on (\ref{tau_j})\;
        \textbf{Compute:}
        $p{(s+1)}=\Bigr[ \hat{p}(s) \Bigr]_{\tau}^{p^{\rm MAX}}$\hspace*{1.75cm}(\ref{power_update})\;
    
    $w = \left[\frac{1}{p_1 + \delta },\dots,\frac{1}{p_L + \delta }\right]$\;
    \tcp{$\delta$ small constant to avoid numerical instability}
    \textbf{Compute:} $f\left(X(s),\varepsilon,p(s)\right)$ \;
    $s = s+1$\;
}
\Return{X,$\varepsilon$,p}\;
\end{algorithm}
\section{Simulation results and analysis}
\label{sec:Simulation_Results}

In this section, we evaluate the performance of our framework within the context of an integrated TN-NTN for a duration of $24$ hours, with a varying number of UEs at each hour, as illustrated in \cite{Chen_2011}. We consider a rural scenario where the terrestrial MBSs are deployed in a hexagonal grid pattern \cite{3GPPTR36.942}.
We focus our study on an area of roughly $2500$ $\mathrm{km}^2$, which corresponds to the coverage range of an LEO satellite beam \cite{3GPPTR38.821}.
Moreover, we assume that the LEO constellation employs earth-fixed beams \cite{3GPPTR38.811} and that the beam serving our scenario originates from a satellite with a $90$ degree elevation angle.
All of the UEs are distributed uniformly throughout the grid.
We provide two settings as benchmarks: The \texttt{3GPP-TN} setting in which there is no satellite tier and the terrestrial tier gets a total bandwidth of $10$ MHz, and the \texttt{3GPP-NTN} setting, where the bandwidth is split according to the 3GPP specifications \cite{3GPPTR38.821}, meaning that $30$ MHz is allocated to the satellite tier and $10$ MHz to the terrestrial tier.
In both settings, the UEs associate with the BSs following the max-RSRP rule and there is no DL transmit power optimization or MBS shutdown.
Note that $\lambda$ is set such that it is inversely proportional to the number of UEs in the network.
Most of the relevant simulation parameters can be found in Table 
\ref{simul_params}, set accordingly to \cite{3GPPTR36.763, 3GPPTR38.811, 3GPPTR38.821,3GPPTR38.901, 3GPPTR36.814, 3GPPTR36.931}.
\begin{table}[h!]
\begin{center}
\begin{tabular}{|l|l|}
\hline Parameter & Value \\
\hline Total Bandwidth $W$  & $40$ $\mathrm{MHz}$ \\
\hline Carrier frequency $f_c$  & $2$ $\mathrm{GHz}$ \\
\hline Subcarrier Spacing   & $15$ $\mathrm{kHz}$ \\
\hline Urban/Rural Inter-Site Distance & $500/1732$ $m$ \\
\hline Number of Macro BSs & $1067$ \\
\hline Satellite Altitude \cite{3GPPTR38.821} & $600$ km \\
\hline Terrestrial Max Tx Power per RE $p_{j}^{MAX}$ \cite{3GPPTR36.814} & $ 17.7 \text{ } \mathrm{dBm}$ \\
\hline Satellite Max Tx Power per RE $p_{j}^{MAX}$ \cite{3GPPTR38.821} & $15.8 \text{ } \mathrm{dBm}$ \\
\hline Antenna gain (Terrestrial) $G_{T_X}$ \cite{3GPPTR36.931} & $14\text{ } \mathrm{dBi}$ \\
\hline Antenna gain (Satellite) $G_{T_X}$ \cite{3GPPTR38.821} & $30\text{ } \mathrm{dBi}$ \\
\hline Shadowing Loss (Terrestrial) $SF$ \cite{3GPPTR38.901} & $4\text{ } - \text{ } 8\text{ } \mathrm{dB}$ \\
\hline Shadowing Loss (Satellite) $SF$ \cite{3GPPTR38.811} & $0\text{ } - \text{ } 12\text{ } \mathrm{dB}$ \\
\hline Line-of-Sight Probability (Satellite / Terrestrial) & Refer to \cite{3GPPTR38.811} / \cite{3GPPTR38.901} \\
\hline White Noise Power Density & $-174 \text{ }$ $ \mathrm{dBm} /  \mathrm{Hz}$ \\
\hline Coverage threshold $RSRP_{\rm min}$ & $-120 \text{ }$ $ \mathrm{dBm}$ \\
\hline
\end{tabular}

\end{center}
\caption{Simulation parameters.}
\label{simul_params}
\end{table}

\subsection{Sum Throughput Analysis}\label{SubSeq_ST_Analysis}
\begin{figure}[h]
    \centering
    \includegraphics[width=  \linewidth ]{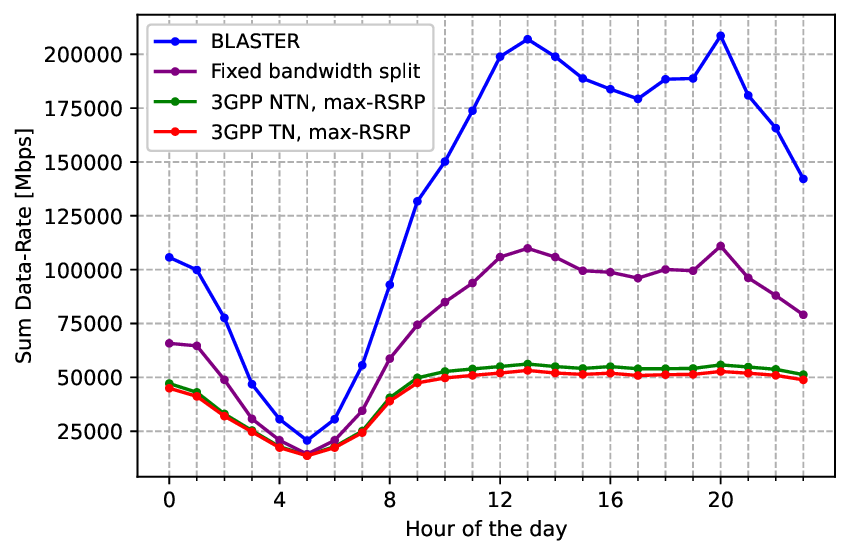}
    \caption{Network Sum Throughput throughout the day for different scenarios.} 
    \label{SLT_plot}
    \vspace{-0.2cm}
\end{figure}

In this section, we analyze the sum throughput of the network, to better understand how our framework adapts the offered capacity to the actual load of the network. 
In this regard, Fig. \ref{SLT_plot} shows the evolution of the sum throughput (ST) during an entire day and for different traffic states.
On top of the two benchmarks, we plot the hourly ST for 
1) BLASTER (presented in Sec. \ref{sec:UtilityOptimization}) and 
2) a setting where the association and power control are based on our algorithm, but bandwidth is shared equally among both tiers (\texttt{Fixed bandwidth split}).
Firstly, we notice that although the network ST gain provided by our framework is limited during low and average traffic periods (e.g. early morning), we see a clear improvement during high-traffic hours.
This is explained by the fact that when the traffic load is low, the algorithm favours energy saving since the value of $\lambda$ has been set to a greater value, to penalize power consumption and drive the terrestrial MBSs to shut down at the expense of the ST.
Inversely, when traffic is high, our framework distributes the load and redistributes the bandwidth resources so that the network ST improves notably.
Indeed, compared to the \texttt{3GPP-NTN} benchmark, we see a growth of the sum throughput which can go up to as much as $270 \%$ during peak traffic.
As expected, the \texttt{3GPP-TN} setting provides the worst performance amongst the compared solutions. This is due to the fact that, without satellites, roughly $7\%$ of the UEs are out of coverage \cite{Alam_2023_2} and the overall available total bandwidth is limited.
\texttt{3GPP-NTN} improves the ST due to the addition of a satellite tier, which provides coverage for the entire grid, as well as a solution for UEs who are associated to overloaded terrestrial MBSs and perceiving a low throughput.
Moreover, considering the settings with our power optimization and UE-BS association, we see that the ST further improves.
This results from reduced interference from neighbouring terrestrial MBSs due to the power control step, leading to a throughput increase.
Furthermore, we notice that the dynamic allocation of the bandwidth benefits the network ST, as we can see it double compared to \texttt{Fixed Bandwidth split} during high traffic.
Finally, we observe an average increase of $249 \%$ in terms of mean throughput during high traffic compared to \texttt{3GPP-NTN}, which underlines the effectiveness of our framework.

\subsection{Satellite Offloading Analysis}\label{Subseq_sat_prop}
\begin{figure}[h]
    \centering
    \includegraphics[width= 0.45\textwidth]{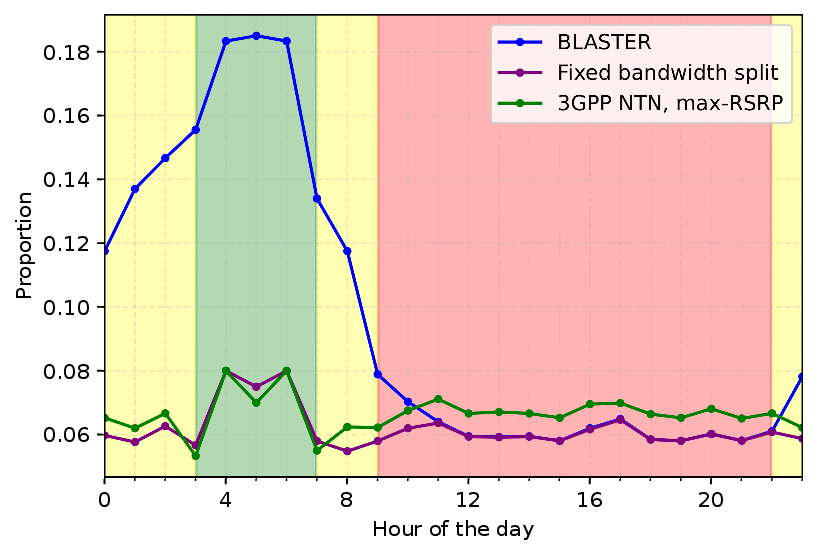}
    \caption{Proportion of UEs associated to the satellite throughout the day.} 
    \label{prop_sat_hourly}
\end{figure}

In this section, we study the critical role that the satellite plays in our framework by analyzing the proportion (see Fig. \ref{prop_sat_hourly}) of UEs associated to a satellite throughout the day.
\newline
The load of the traffic is represented by the background colour, i.e. green, yellow, and red are low, average, and high-traffic hours respectively.
We know that during low traffic, the onus is on reducing the TN power consumption. Therefore, the satellite becomes an attractive means to serve UEs.
Thanks to our method, this results in an increased proportion (more than a $200 \%$ increase) of UEs associated to a satellite compared to the benchmark \texttt{3GPP-NTN}.
As a result, we are able to shut down more terrestrial MBSs during low-traffic hours.
On the contrary, we see that the satellite associates with fewer UEs during high-traffic hours. Indeed, the proportion of UEs associated to it is lesser than in the \texttt{3GPP-NTN} scenario. This can be explained by the fact that when the traffic is high, the satellite will act more as a coverage layer.
Most of the satellite bandwidth is allocated to the terrestrial MBSs, as the TN can support higher throughputs due to the large spectrum reuse.
Therefore, we notice how the algorithm is able to adapt the overall TN-NTN resources according to traffic variations.
\subsection{Power Consumption Analysis}
\begin{figure}[h]
    \centering
\includegraphics[width = 0.45 \textwidth]{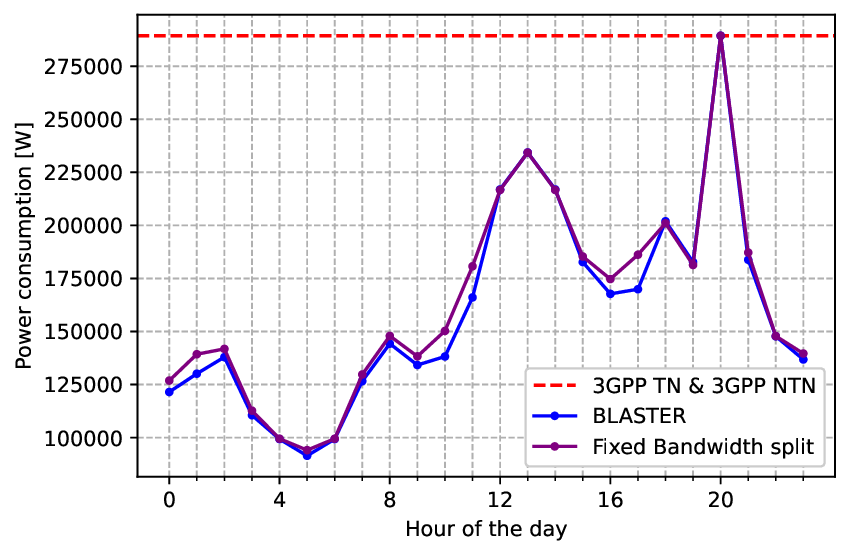}
    \caption{Network Power consumption throughout the day.} 
    \label{power_conso}
\end{figure}

In this section, we investigate the performance of our framework in terms of power consumption.
As mentioned above, terrestrial MBSs transmit power and activation control plays an important role in our proposed framework.
In Fig. \ref{power_conso}, we show the network power consumption throughout the day for the various solutions analyzed in this paper.
The dotted red line represents the power consumption level for both \texttt{3GPP} benchmarks. As explained before, since there is no power optimization or terrestrial MBS shutdown in those settings, the transmit power levels are at the maximum throughout the day.
We observe an average decrease of the network power consumption by $65.4 \%$ during low traffic for our framework, compared to both \texttt{3GPP} benchmarks.
This is because, as we discussed in Sec. \ref{Subseq_sat_prop}, the satellites take a prominent role at this period of the day, associating with more UEs than in the \texttt{3GPP-NTN} network setting. This allows the shutdown of many terrestrial MBSs and saves a lot of energy in the process.
Also, during high traffic, we notice that the average network power consumption decreases by $33 \%$. As the shutdown of MBSs is not the priority during this period, the energy saved is mostly due to some BSs decreasing their transmit power without negatively impacting the QoS.
The noticeable improvement in ST observed in Fig. \ref{SLT_plot} during high traffic corroborates this statement.
Once again, this highlights the ability of our network to adapt to the demands imposed by the traffic state, with the satellite playing an eminent role throughout.

\section{Conclusion}
\label{sec:conclusions}
In this paper, we have introduced BLASTER, a framework that operates in an integrated TN-NTN, and is able to manage UE association, regulate power, and allocate bandwidth between the terrestrial and non-terrestrial tiers of the network.
Our proposal is to adapt the network behaviour according to the traffic demand, i.e. focus on energy saving during low-traffic hours and offer large throughput otherwise.
Simulation results show that BLASTER can greatly improve the network, enhancing the mean throughput during high-traffic hours, and scaling down the power consumption during low-traffic hours.
We also underline the crucial and evolving role that the NTNs play in the success of our framework, emphasizing the decisive part they are poised to play in ensuring ubiquitous and sustainable mobile services in the coming years.


\bibliographystyle{IEEEtran}
\bibliography{reference.bib}

\begin{acronym}[AAAAAAAAA]

 \acro{2D}{two-dimensional}
 \acro{3D}{three-dimensional}
 \acro{3G}{third generation}
 \acro{3GPP}{third generation partnership project}
 \acro{4G}{fourth generation}
 \acro{5G}{fifth generation}
 \acro{5GC}{5G Core Network}
 \acro{AAA}{authentication, authorisation and accounting}
 \acro{ABS}{almost blank subframe}
 \acro{AC}{alternating current}
 \acro{ACIR}{adjacent channel interference rejection ratio}
 \acro{ACK}{acknowledgment}
 \acro{ACL}{allowed CSG list}
 \acro{ACLR}{adjacent channel leakage ratio}
 \acro{ACPR}{adjacent channel power ratio}
 \acro{ACS}{adjacent channel selectivity}
 \acro{ADC}{analog-to-digital converter}
 \acro{ADSL}{asymmetric digital subscriber line}
 \acro{AEE}{area energy efficiency}
 \acro{AF}{amplify-and-forward}
 \acro{AGCH}{access grant channel}
 \acro{AGG}{aggressor cell}
 \acro{AH}{authentication header}
 \acro{AI}{artificial intelligence}
 \acro{AKA}{authentication and key agreement}
 \acro{AMC}{adaptive modulation and coding}
 \acro{ANN}{artificial neural network}
 \acro{ANR}{automatic neighbour relation}
 \acro{AoA}{angle of arrival}
 \acro{AoD}{angle of departure}
 \acro{APC}{area power consumption}
 \acro{API}{application programming interface}
 \acro{APP}{a posteriori probability}
 \acro{AR}{augmented reality}
 \acro{ARIMA}{autoregressive integrated moving average}
 \acro{ARQ}{automatic repeat request}
 \acro{AS}{access stratum}
 \acro{ASE}{area spectral efficiency}
 \acro{ASM}{advanced sleep mode}
 \acro{ASN}{access service network}
 \acro{ATM}{asynchronous transfer mode}
 \acro{ATSC}{Advanced Television Systems Committee}
 \acro{AUC}{authentication centre}
 \acro{AWGN}{additive white gaussian noise}
 \acro{BB}{baseband}
 \acro{BBU}{baseband unit}
 \acro{BCCH}{broadcast control channel}
 \acro{BCH}{broadcast channel}
 \acro{BCJR}{Bahl-Cocke-Jelinek-Raviv} 
 \acro{BE}{best effort}
 \acro{BER}{bit error rate}
 \acro{BLER}{block error rate}
 \acro{BPSK}{binary phase-shift keying}
 \acro{BR}{bit rate}
 \acro{BS}{base station}
 \acro{BSC}{base station controller}
 \acro{BSIC}{base station identity code}
 \acro{BSP}{binary space partitioning}
 \acro{BSS}{blind source separation}
 \acro{BTS}{base transceiver station}
 \acro{BWP}{Bandwidth Part}
 \acro{CA}{carrier aggregation}
 \acro{CAC}{call admission control}
 \acro{CaCo}{carrier component}
 \acro{CAPEX}{capital expenditure}
 \acro{capex}{capital expenses}
 \acro{CAS}{cluster angular spread}
 \acro{CATV}{community antenna television}
 \acro{CAZAC}{constant amplitude zero auto-correlation}
 \acro{CC}{component carrier}
 \acro{CCCH}{common control channel}
 \acro{CCDF}{complementary cumulative distribution function}
 \acro{CCE}{control channel element}
 \acro{CCO}{coverage and capacity optimisation}
 \acro{CCPCH}{common control physical channel}
 \acro{CCRS}{coordinated and cooperative relay system}
 \acro{CCTrCH}{coded composite transport channel}
 \acro{CDF}{cumulative distribution function}
 \acro{CDMA}{code division multiple access}
 \acro{CDS}{cluster delay spread}
 \acro{CESM}{capacity effective SINR mapping}
 \acro{CO$_{2e}$}{carbon dioxide equivalent}
 \acro{CFI}{control format indicator}
 \acro{CFL}{Courant-Friedrichs-Lewy}
 \acro{CGI}{cell global identity}
 \acro{CID}{connection identifier}
 \acro{CIF}{carrier indicator field}
 \acro{CIO}{cell individual offset}
 \acro{CIR}{channel impulse response}
 \acro{CNN}{Convolutional Neural Network}
 \acro{CMF}{cumulative mass function}
 \acro{C-MIMO}{cooperative MIMO}
 \acro{CN}{core network}
 \acro{COC}{cell outage compensation}
 \acro{COD}{cell outage detection}
 \acro{CoMP}{coordinated multi-point}
 \acro{ConvLSTM}{Convolutional LSTM}
 \acro{CP}{cycle prefix}
 \acro{CPC}{cognitive pilot channel}
 \acro{CPCH}{common packet channel}
 \acro{CPE}{customer premises equipment}
 \acro{CPICH}{common pilot channel}
 \acro{CPRI}{common public radio interface}
 \acro{CPU}{central processing unit}
 \acro{CQI}{channel quality indicator}
 \acro{CR}{cognitive radio}
 \acro{CRAN}{centralized radio access network} 
 \acro{C-RAN}{cloud radio access network} 
 \acro{CRC}{cyclic redundancy check}
 \acro{CRE}{cell range expansion}
 \acro{C-RNTI}{cell radio network temporary identifier}
 \acro{CRP}{cell re-selection parameter}
 \acro{CRS}{cell-specific reference symbol}
 \acro{CRT}{cell re-selection threshold}
 \acro{CSCC}{common spectrum coordination channel}
 \acro{CSG ID}{closed subscriber group ID}
 \acro{CSG}{closed subscriber group}
 \acro{CSI}{channel state information}
 \acro{CSIR}{receiver-side channel state information}
 \acro{CSI-RS}{channel state information-reference signals}
 \acro{CSO}{cell selection offset}
 \acro{CTCH}{common traffic channel}
 \acro{CTS}{clear-to-send} 
 \acro{CU}{central unit}
 \acro{CV}{cross-validation}
 \acro{CWiND}{Centre for Wireless Network Design}
 \acro{D2D}{device to device}
 \acro{DAB}{digital audio broadcasting}
 \acro{DAC}{digital-to-analog converter}
 \acro{DAS}{distributed antenna system}
 \acro{dB}{decibel}
 \acro{dBi}{isotropic-decibel}
 \acro{DC}{direct current}
 \acro{DCCH}{dedicated control channel}
 \acro{DCF}{decode-and-forward}
 \acro{DCH}{dedicated channel}
 \acro{DC-HSPA}{dual-carrier high speed packet access}
 \acro{DCI}{downlink control information}
 \acro{DCM}{directional channel model}
 \acro{DCP}{dirty-paper coding}
 \acro{DCS}{digital communication system}
 \acro{DECT}{digital enhanced cordless telecommunication}
 \acro{DeNB}{donor eNodeB}
 \acro{DFP}{dynamic frequency planning}
 \acro{DFS}{dynamic frequency selection}
 \acro{DFT}{discrete Fourier transform}
 \acro{DFTS}{discrete Fourier transform spread}
 \acro{DHCP}{dynamic host control protocol}
 \acro{DL}{downlink}
 \acro{DMC}{dense multi-path components}
 \acro{DMF}{demodulate-and-forward}
 \acro{DMT}{diversity and multiplexing tradeoff}
  \acro{DNN}{deep neural network} 
 \acro{DoA}{direction-of-arrival}
 \acro{DoD}{direction-of-departure}
 \acro{DoS}{denial of service}
 \acro{DPCCH}{dedicated physical control channel}
 \acro{DPDCH}{dedicated physical data channel}
 \acro{D-QDCR}{distributed QoS-based dynamic channel reservation}
 \acro{DQL}{deep Q-learning}
  \acro{DRAN}{distributed radio access network}
 \acro{DRS}{discovery reference signal}
 \acro{DRL}{deep reinforcement learning}
 \acro{DRX}{discontinuous reception}
 \acro{DS}{down stream}
 \acro{DSA}{dynamic spectrum access}
 \acro{DSCH}{downlink shared channel}
 \acro{DSL}{digital subscriber line}
 \acro{DSLAM}{digital subscriber line access multiplexer}
 \acro{DSP}{digital signal processor}
 \acro{DT}{decision tree}
 \acro{DTCH}{dedicated traffic channel}
 \acro{DTX}{discontinuous transmission}
   \acro{DU}{distributed unit}
 \acro{DVB}{digital video broadcasting}
 \acro{DXF}{drawing interchange format}
 \acro{E2E}{end-to-end}
 \acro{EAGCH}{enhanced uplink absolute grant channel}
 \acro{EA}{evolutionary algorithm}
 \acro{EAP}{extensible authentication protocol}
 \acro{EC}{evolutionary computing}
 \acro{ECGI}{evolved cell global identifier}
 \acro{ECR}{energy consumption ratio}
 \acro{ECRM}{effective code rate map}
 \acro{EDCH}{enhanced dedicated channel}
 \acro{EE}{energy efficiency}
 \acro{EESM}{exponential effective SINR mapping}
 \acro{EF}{estimate-and-forward}
 \acro{EGC}{equal gain combining}
 \acro{EHICH}{EDCH HARQ indicator channel}
 \acro{eICIC}{enhanced intercell interference coordination}
 \acro{EIR}{equipment identity register}
 \acro{EIRP}{effective isotropic radiated power}
 \acro{ELF}{evolutionary learning of fuzzy rules}
 \acro{eMBB}{enhanced mobile broadband}
  \acro{EMR}{Electromagnetic-Radiation}
 \acro{EMS}{enhanced messaging service}
 \acro{eNB}{evolved NodeB}
 \acro{eNodeB}{evolved NodeB}
 \acro{EoA}{elevation of arrival}
 \acro{EoD}{elevation of departure}
 \acro{EPB}{equal path-loss boundary}
 \acro{EPC}{evolved packet core}
 \acro{EPDCCH}{enhanced physical downlink control channel}
 \acro{EPLMN}{equivalent PLMN}
 \acro{EPS}{evolved packet system}
 \acro{ERAB}{eUTRAN radio access bearer}
 \acro{ERGC}{enhanced uplink relative grant channel}
 \acro{ERTPS}{extended real time polling service}
 \acro{ESB}{equal downlink receive signal strength boundary}
 \acro{ESF}{even subframe}
 \acro{ESP}{encapsulating security payload}
 \acro{ETSI}{European Telecommunications Standards Institute}
 \acro{E-UTRA}{evolved UTRA}
 \acro{EU}{European Union}
 \acro{EUTRAN}{evolved UTRAN}
 \acro{EVDO}{evolution-data optimised}
 \acro{FACCH}{fast associated control channel}
 \acro{FACH}{forward access channel}
 \acro{FAP}{femtocell access point}
 \acro{FARL}{fuzzy assisted reinforcement learning}
 \acro{FCC}{Federal Communications Commission}
 \acro{FCCH}{frequency-correlation channel}
 \acro{FCFS}{first-come first-served}
 \acro{FCH}{frame control header}
 \acro{FCI}{failure cell ID}
 \acro{FD}{frequency-domain}
 \acro{FDD}{frequency division duplexing}
 \acro{FDM}{frequency division multiplexing}
 \acro{FDMA}{frequency division multiple access}
 \acro{FDTD}{finite-difference time-domain}
 \acro{FE}{front-end}
 \acro{FeMBMS}{further evolved multimedia broadcast multicast service}
 \acro{FER}{frame error rate}
 \acro{FFR}{fractional frequency reuse}
 \acro{FFRS}{fractional frequency reuse scheme}
 \acro{FFT}{fast Fourier transform}
 \acro{FFU}{flexible frequency usage}
 \acro{FGW}{femtocell gateway}
 \acro{FIFO}{first-in first-out}
 \acro{FIS}{fuzzy inference system}
 \acro{FMC}{fixed mobile convergence}
 \acro{FPC}{fractional power control}
 \acro{FPGA}{field-programmable gate array}
 \acro{FRS}{frequency reuse scheme}
 \acro{FTP}{file transfer protocol}
 \acro{FTTx}{fiber to the x}
 \acro{FUSC}{full usage of subchannels}
 \acro{GA}{genetic algorithm}
 \acro{GAN} {generic access network}
 \acro{GANC}{generic access network controller}
 \acro{GBR}{guaranteed bitrate}
 \acro{GCI}{global cell identity}
 \acro{STGCN}{Spatio-Temporal Graph convolutional network}
 \acro{GERAN}{GSM edge radio access network}
 \acro{GGSN}{gateway GPRS support node}
 \acro{GHG}{greenhouse gas}
 \acro{GMSC}{gateway mobile switching centre}
 \acro{gNB}{next generation NodeB}
 \acro{GNN}{Graph Neural Network}
 \acro{GNSS}{global navigation satellite system}
 \acro{GP}{genetic programming}
 \acro{GPON}{Gigabit passive optical network}
 \acro{GPP}{general purpose processor}
 \acro{GPRS}{general packet radio service}
 \acro{GPS}{global positioning system}
 \acro{GPU}{graphics processing unit}
 \acro{GRU}{gated recurrent unit}
 \acro{GSCM}{geometry-based stochastic channel models}
 \acro{GSM}{global system for mobile communication}
 \acro{GTD}{geometry theory of diffraction}
 \acro{GTP}{GPRS tunnel protocol}
 \acro{GTP-U}{GPRS tunnel protocol - user plane}
 \acro{HA}{historical average}
 \acro{HARQ}{hybrid automatic repeat request}
 \acro{HBS}{home base station}
 \acro{HCN}{heterogeneous cellular network}
 \acro{HCS}{hierarchical cell structure}
  \acro{HD}{high definition}
 \acro{HDFP}{horizontal dynamic frequency planning}
 \acro{HeNB}{home eNodeB}
 \acro{HeNodeB}{home eNodeB}
 \acro{HetNet}{heterogeneous network}
 \acro{HiFi}{high fidelity}
 \acro{HII}{high interference indicator}
 \acro{HLR}{home location register}
 \acro{HNB}{home NodeB}
 \acro{HNBAP}{home NodeB application protocol}
 \acro{HNBGW}{home NodeB gateway}
 \acro{HNodeB}{home NodeB}
 \acro{HO}{handover}
 \acro{HOF}{handover failure}
 \acro{HOM}{handover hysteresis margin}
 \acro{HPBW}{half power beam width}
 \acro{HPLMN}{home PLMN}
 \acro{HPPP}{homogeneous Poison point process}
 \acro{HRD}{horizontal reflection diffraction}
 \acro{HSB}{hot spot boundary}
 \acro{HSDPA}{high speed downlink packet access}
 \acro{HSDSCH}{high-speed DSCH}
 \acro{HSPA}{high speed packet access}
 \acro{HSS}{home subscriber server}
 \acro{HSUPA}{high speed uplink packet access}
 \acro{HUA}{home user agent}
 \acro{HUE}{home user equipment}
 \acro{HVAC}{heating, ventilating, and air conditioning}
 \acro{HW}{Holt-Winters}
 \acro{IC}{interference cancellation}
 \acro{ICI}{inter-carrier interference}
 \acro{ICIC}{intercell interference coordination}
 \acro{ICNIRP}{International Commission on Non-Ionising Radiation Protection}
 \acro{ICS}{IMS centralised service}
 \acro{ICT}{information and communication technology}
 \acro{ID}{identifier}
 \acro{IDFT}{inverse discrete Fourier transform}
 \acro{IE}{information element}
 \acro{IEEE}{Institute of Electrical and Electronics Engineers}
 \acro{IETF}{Internet engineering task force}
 \acro{IFA}{Inverted-F-antennas}
 \acro{IFFT}{inverse fast Fourier transform}
 \acro{i.i.d.}{independent and identical distributed}
 \acro{IIR}{infinite impulse response}
 \acro{IKE}{Internet key exchange}
 \acro{IKEv2}{Internet key exchange version 2}
 \acro{ILP}{integer linear programming}
 \acro{IMEI}{international mobile equipment identity}
 \acro{IMS}{IP multimedia subsystem}
 \acro{IMSI}{international mobile subscriber identity}
 \acro{IMT}{international mobile telecommunications}
 \acro{INH}{indoor hotspot}
 \acro{IOI}{interference overload indicator}
 \acro{IoT}{Internet of things}
 \acro{IP}{Internet protocol}
 \acro{IPSEC}{Internet protocol security}
 \acro{IR}{incremental redundancy}
 \acro{IRC}{interference rejection combining}
 \acro{ISD}{inter site distance}
 \acro{ISI}{inter symbol interference}
 \acro{ITU}{International Telecommunication Union}
 \acro{Iub}{UMTS interface between RNC and NodeB}
 \acro{IWF}{IMS interworking function}
 \acro{JFI}{Jain's fairness index}
 \acro{KPI}{key performance indicator}
 \acro{KNN}{k-nearest neighbours}
 \acro{L1}{layer one}
 \acro{L2}{layer two}
 \acro{L3}{layer three}
 \acro{LA}{location area}
 \acro{LAA}{licensed Assisted Access}
 \acro{LAC}{location area code}
 \acro{LAI}{location area identity}
 \acro{LAU}{location area update}
 \acro{LDA}{linear discriminant analysis} 
 \acro{LIDAR}{laser imaging detection and ranging}
 \acro{LLR}{log-likelihood ratio}
 \acro{LLS}{link-level simulation}
 \acro{LMDS}{local multipoint distribution service}
 \acro{LMMSE}{linear minimum mean-square-error}
 \acro{LoS}{line-of-sight}
 \acro{LPC}{logical PDCCH candidate}
 \acro{LPN}{low power node}
 \acro{LR}{likelihood ratio}
 \acro{LSAS}{large-scale antenna system}
 \acro{LSP}{large-scale parameter}
 \acro{LSTM}{long short term memory cell}
 \acro{LTE/SAE}{long term evolution/system architecture evolution}
 \acro{LTE}{long term evolution}
 \acro{LTE-A}{long term evolution advanced}
 \acro{LUT}{look up table}
 \acro{MAC}{medium access control}
 \acro{MaCe}{macro cell}
  \acro{MAE}{mean absolute error}
 \acro{MAP}{media access protocol}
 \acro{MAPE}{mean absolute percentage error}
 \acro{MAXI}{maximum insertion}
 \acro{MAXR}{maximum removal}
 \acro{MBMS}{multicast broadcast multimedia service} 
 \acro{MBS}{macrocell base station}
 \acro{MBSFN}{multicast-broadcast single-frequency network}
 \acro{MC}{modulation and coding}
 \acro{MCB}{main circuit board}
 \acro{MCM}{multi-carrier modulation}
 \acro{MCP}{multi-cell processing}
 \acro{MCS}{modulation and coding scheme}
 \acro{MCSR}{multi-carrier soft reuse}
 \acro{MDAF}{management data analytics function}
 \acro{MDP}{markov decision process }
 \acro{MDT}{minimisation of drive tests}
 \acro{MEA}{multi-element antenna}
 \acro{MeNodeB}{Master eNodeB}
 \acro{MGW}{media gateway}
 \acro{MIB}{master information block}
 \acro{MIC}{mean instantaneous capacity}
 \acro{MIESM}{mutual information effective SINR mapping}
 \acro{MIMO}{multiple-input multiple-output}
 \acro{MINI}{minimum insertion}
 \acro{MINR}{minimum removal}
 \acro{MIP}{mixed integer program}
 \acro{MISO}{multiple-input single-output}
 \acro{ML}{machine learning}
 \acro{MLB}{mobility load balancing}
 \acro{MLB}{mobility load balancing}
 \acro{MM}{mobility management}
 \acro{MME}{mobility management entity}
 \acro{mMIMO}{massive multiple-input multiple-output}
 \acro{MMSE}{minimum mean square error}
 \acro{mMTC}{massive machine type communication}
 \acro{MNC}{mobile network code}
 \acro{MNO}{mobile network operator}
 \acro{MOS}{mean opinion score}
 \acro{MPC}{multi-path component}
 \acro{MR}{measurement report}
 \acro{MRC}{maximal ratio combining}
 \acro{MR-FDPF}{multi-resolution frequency-domain parflow}
 \acro{MRO}{mobility robustness optimisation}
 \acro{MRT}{Maximum Ratio Transmission}
 \acro{MS}{mobile station}
 \acro{MSC}{mobile switching centre}
 \acro{MSE}{mean square error}
 \acro{MSISDN}{mobile subscriber integrated services digital network number}
 \acro{MUE}{macrocell user equipment}
 \acro{MU-MIMO}{multi-user MIMO}
 \acro{MVNO}{mobile virtual network operators}
 \acro{NACK}{negative acknowledgment}
 \acro{NAS}{non access stratum}
 \acro{NAV}{network allocation vector}
 \acro{NB}{Naive Bayes}   
 \acro{NCL}{neighbour cell list}
 \acro{NEE}{network energy efficiency}
  \acro{NF}{Network Function}
 \acro{NFV}{Network Functions Virtualization}
 \acro{NG}{next generation}
 \acro{NGMN}{next generation mobile networks}
 \acro{NG-RAN}{next generation radio access network} 
 \acro{NIR}{non ionisation radiation}
 \acro{NLoS}{non-line-of-sight}
 \acro{NN}{nearest neighbour} 
 \acro{NR}{new radio}
 \acro{NRTPS}{non-real-time polling service}
 \acro{NSS}{network switching subsystem}
 \acro{NTP}{network time protocol}
 \acro{NWG}{network working group}
 \acro{NWDAF}{network data analytics function} 
 \acro{OA}{open access}
 \acro{OAM}{operation, administration and maintenance}
 \acro{OC}{optimum combining}
 \acro{OCXO}{oven controlled oscillator}
 \acro{ODA}{omdi-directional antenna} 
 \acro{ODU}{optical distribution unit}
 \acro{OFDM}{orthogonal frequency division multiplexing}
 \acro{OFDMA}{orthogonal frequency division multiple access}
 \acro{OFS}{orthogonally-filled subframe}
 \acro{OLT}{optical line termination}
 \acro{ONT}{optical network terminal}
 \acro{OPEX}{operational expenditure}
 \acro{OSF}{odd subframe}
 \acro{OSI}{open systems interconnection}
 \acro{OSS}{operation support subsystem}
 \acro{OTT}{over the top}
 \acro{P2MP}{point to multi-point}
 \acro{P2P}{point to point}
 \acro{PAPR}{peak-to-average power ratio}
 \acro{PA}{power amplifier}
 \acro{PBCH}{physical broadcast channel}
 \acro{PC}{power control}
 \acro{PCB}{printed circuit board}
 \acro{PCC}{primary carrier component}
 \acro{PCCH}{paging control channel}
 \acro{PCCPCH}{primary common control physical channel}
 \acro{PCell}{primary cell}
 \acro{PCFICH}{physical control format indicator channel}
 \acro{PCH}{paging channel}
 \acro{PCI}{physical layer cell identity}
 \acro{PCPICH}{primary common pilot channel}
 \acro{PCPPH}{physical common packet channel}
 \acro{PDCCH}{physical downlink control channel}
 \acro{PDCP}{packet data convergence protocol}
 \acro{PDF}{probability density function}
 \acro{PDSCH}{physical downlink shared channel}
 \acro{PDU}{packet data unit}
 \acro{PeNB}{pico eNodeB}
 \acro{PeNodeB}{pico eNodeB}
 \acro{PF}{proportional fair}
 \acro{PGW}{packet data network gateway}
 \acro{PGFL}{probability generating functional}
 \acro{PhD}{doctor of philosophy}
 \acro{PHICH}{physical HARQ indicator channel}
 \acro{PHY}{physical}
 \acro{PIC}{parallel interference cancellation}
 \acro{PKI}{public key infrastructure}
 \acro{PL}{path loss}
 \acro{PMI}{precoding-matrix indicator}
 \acro{PLMN ID}{public land mobile network identity}
 \acro{PLMN}{public land mobile network}
 \acro{PML}{perfectly matched layer}
 \acro{PMF}{probability mass function}
 \acro{PMP}{point to multi-point}
 \acro{PN}{pseudorandom noise}
 \acro{POI}{point of interest}
 \acro{PON}{passive optical network}
 \acro{POP}{point of presence}
 \acro{PP}{point process}
 \acro{PPP}{Poisson point process}
 \acro{PPT}{PCI planning tools}
 \acro{PRACH}{physical random access channel}
 \acro{PRB}{physical resource block}
 \acro{PSC}{primary scrambling code}
 \acro{PSD}{power spectral density}
 \acro{PSS}{primary synchronisation channel}
 \acro{PSTN}{public switched telephone network}
 \acro{PTP}{point to point}
 \acro{PUCCH}{Physical Uplink Control Channel}
 \acro{PUE}{picocell user equipment}
 \acro{PUSC}{partial usage of subchannels}
 \acro{PUSCH}{physical uplink shared channel}
 \acro{QAM}{quadrature amplitude modulation}
 \acro{QCI}{QoS class identifier}
 \acro{QoE}{quality of experience}
 \acro{QoS}{quality of service}
 \acro{QPSK}{quadrature phase-shift keying}
 \acro{RAB}{radio access bearer}
 \acro{RACH}{random access channel}
 \acro{RADIUS}{remote authentication dial-in user services}
 \acro{RAN}{radio access network}
 \acro{RANAP}{radio access network application part}
 \acro{RAT}{radio access technology}
 \acro{RAU}{remote antenna unit}
 \acro{RAXN}{relay-aided x network}
 \acro{RB}{resource block}
 \acro{RCI}{re-establish cell id}
 \acro{RE}{resource efficiency}
 \acro{REB}{range expansion bias}
 \acro{REG}{resource element group}
 \acro{RF}{radio frequency}
  \acro{RFID}{radio frequency identification}
 \acro{RFP}{radio frequency planning}
 \acro{RI}{rank indicator}
 \acro{RL}{reinforcement learning}
 \acro{RLC}{radio link control}
 \acro{RLF}{radio link failure}
 \acro{RLM}{radio link monitoring}
 \acro{RMA}{rural macrocell}
 \acro{RMS}{root mean square}
 \acro{RMSE}{root mean square error}
 \acro{RN}{relay node}
 \acro{RNC}{radio network controller}
 \acro{RNL}{radio network layer}
 \acro{RNN}{recurrent neural network}
 \acro{RNP}{radio network planning}
 \acro{RNS}{radio network subsystem}
 \acro{RNTI}{radio network temporary identifier}
 \acro{RNTP}{relative narrowband transmit power}
 \acro{RPLMN}{registered PLMN}
 \acro{RPSF}{reduced-power subframes}
 \acro{RR}{round robin}
 \acro{RRC}{radio resource control}
 \acro{RRH}{remote radio head}
 \acro{RRM}{radio resource management}
 \acro{RS}{reference signal}
 \acro{RSC}{recursive systematic convolutional}
 \acro{RS-CS}{resource-specific cell-selection}
 \acro{RSQ}{reference signal quality}
 \acro{RSRP}{reference signal received power}
 \acro{RSRQ}{reference signal received quality}
 \acro{RSS}{reference signal strength}
 \acro{RSSI}{receive signal strength indicator}
 \acro{RTP}{real time transport}
 \acro{RTPS}{real-time polling service}
 \acro{RTS}{request-to-send}
 \acro{RTT}{round trip time}
  \acro{RU}{remote unit}
  \acro{RV}{random variable}
 \acro{RX}{receive}
 \acro{S1-AP}{s1 application protocol}
 \acro{S1-MME}{s1 for the control plane}
 \acro{S1-U}{s1 for the user plane}
 \acro{SA}{simulated annealing}
 \acro{SACCH}{slow associated control channel}
 \acro{SAE}{system architecture evolution}
 \acro{SAEGW}{system architecture evolution gateway}
 \acro{SAIC}{single antenna interference cancellation}
 \acro{SAP}{service access point}
 \acro{SAR}{specific absorption rate}
 \acro{SARIMA}{seasonal autoregressive integrated moving average}
 \acro{SAS}{spectrum allocation server}
 \acro{SBS}{super base station}
 \acro{SCC}{standards coordinating committee}
 \acro{SCCPCH}{secondary common control physical channel}
 \acro{SCell}{secondary cell}
 \acro{SCFDMA}{single carrier FDMA}
 \acro{SCH}{synchronisation channel}
 \acro{SCM}{spatial channel model}
 \acro{SCN}{small cell network}
 \acro{SCOFDM}{single carrier orthogonal frequency division multiplexing}
 \acro{SCP}{single cell processing}
 \acro{SCTP}{stream control transmission protocol}
 \acro{SDCCH}{standalone dedicated control channel}
 \acro{SDMA}{space-division multiple-access}
  \acro{SDO}{standard development organization}
 \acro{SDR}{software defined radio}
 \acro{SDU}{service data unit}
 \acro{SE}{spectral efficiency}
 \acro{SeNodeB}{secondary eNodeB}
 \acro{Seq2Seq}{Sequence-to-sequence}
 \acro{SFBC}{space frequency block coding}
 \acro{SFID}{service flow ID}
 \acro{SG}{signalling gateway}
 \acro{SGSN}{serving GPRS support node}
 \acro{SGW}{serving gateway}
 \acro{SI}{system information}
 \acro{SIB}{system information block}
 \acro{SIB1}{systeminformationblocktype1}
 \acro{SIB4}{systeminformationblocktype4}
 \acro{SIC}{successive interference cancellation}
 \acro{SIGTRAN}{signalling transport}
 \acro{SIM}{subscriber identity module}
 \acro{SIMO}{single input multiple output}
 \acro{SINR}{signal to interference plus noise ratio}
 \acro{SIP}{session initiated protocol}
 \acro{SIR}{signal to interference ratio}
 \acro{SISO}{single input single output}
 \acro{SLAC}{stochastic local area channel}
 \acro{SLL}{secondary lobe level}
 \acro{SLNR}{signal to leakage interference and noise ratio}
 \acro{SLS}{system-level simulation}
 \acro{SMAPE}{symmetric mean absolute percentage error}
 \acro{SMB}{small and medium-sized businesses}
 \acro{SmCe}{small cell}
 \acro{SMS}{short message service}
 \acro{SN}{serial number}
 \acro{SNMP}{simple network management protocol}
 \acro{SNR}{signal to noise ratio}
 \acro{SOCP}{second-order cone programming}
 \acro{SOHO}{small office/home office}
 \acro{SON}{self-organising network}
 \acro{son}{self-organising networks}
 \acro{SOT}{saving of transmissions}
 \acro{SPS}{spectrum policy server}
 \acro{SRS}{sounding reference signals}
 \acro{SS}{synchronization signal}
 \acro{SSL}{secure socket layer}
 \acro{SSMA}{spread spectrum multiple access}
 \acro{SSS}{secondary synchronisation channel}
 \acro{ST}{spatio temporal}
 \acro{STA}{steepest ascent}
 \acro{STBC}{space-time block coding}
 \acro{SUI}{stanford university interim}
 \acro{SVR}{support vector regression}
 \acro{TA}{timing advance}
 \acro{TAC}{tracking area code}
 \acro{TAI}{tracking area identity}
 \acro{TAS}{transmit antenna selection}
 \acro{TAU}{tracking area update}
 \acro{TCH}{traffic channel}
 \acro{TCO}{total cost of ownership}
 \acro{TCP}{transmission control protocol}
 \acro{TCXO}{temperature controlled oscillator}
 \acro{TD}{temporal difference}
 \acro{TDD}{time division duplexing}
 \acro{TDM}{time division multiplexing}
 \acro{TDMA}{time division multiple access}
  \acro{TDoA}{time difference of arrival}
 \acro{TEID}{tunnel endpoint identifier}
 \acro{TLS}{transport layer security}
 \acro{TNL}{transport network layer}
  \acro{ToA}{time of arrival}
 \acro{TP}{throughput}
 \acro{TPC}{transmit power control}
 \acro{TPM}{trusted platform module}
 \acro{TR}{transition region}
 \acro{TS}{tabu search}
 \acro{TSG}{technical specification group}
 \acro{TTG}{transmit/receive transition gap}
 \acro{TTI}{transmission time interval}
 \acro{TTT}{time-to-trigger}
 \acro{TU}{typical urban}
 \acro{TV}{television}
 \acro{TWXN}{two-way exchange network}
 \acro{TX}{transmit}
 \acro{UARFCN}{UTRA absolute radio frequency channel number}
 \acro{UAV}{unmanned aerial vehicle}
 \acro{UCI}{uplink control information}
 \acro{UDP}{user datagram protocol}
 \acro{UDN}{ultra-dense network}
 \acro{UE}{user equipment}
 \acro{UGS}{unsolicited grant service}
 \acro{UICC}{universal integrated circuit card}
 \acro{UK}{united kingdom}
 \acro{UL}{uplink}
 \acro{UMA}{unlicensed mobile access}
 \acro{UMi}{urban micro}
 \acro{UMTS}{universal mobile telecommunication system}
 \acro{UN}{United Nations}
 \acro{URLLC}{ultra-reliable low-latency communication}
 \acro{US}{upstream}
 \acro{USIM}{universal subscriber identity module}
 \acro{UTD}{theory of diffraction}
 \acro{UTRA}{UMTS terrestrial radio access}
 \acro{UTRAN}{UMTS terrestrial radio access network}
 \acro{UWB}{ultra wide band}
 \acro{VD}{vertical diffraction}
 \acro{VDFP}{vertical dynamic frequency planning}
 \acro{VDSL}{very-high-bit-rate digital subscriber line}
 \acro{VeNB}{virtual eNB}
 \acro{VeNodeB}{virtual eNodeB}
 \acro{VIC}{victim cell}
 \acro{VLR}{visitor location register}
 \acro{VNF}{virtual network function}
 \acro{VoIP}{voice over IP}
 \acro{VoLTE}{voice over LTE}
 \acro{VPLMN}{visited PLMN}
 \acro{VR}{visibility region}
  \acro{VRAN}{virtualized radio access network}
 \acro{WCDMA}{wideband code division multiple access}
 \acro{WEP}{wired equivalent privacy}
 \acro{WG}{working group}
 \acro{WHO}{world health organisation}
 \acro{Wi-Fi}{Wi-Fi}
 \acro{WiMAX}{wireless interoperability for microwave access}
 \acro{WiSE}{wireless system engineering}
 \acro{WLAN}{wireless local area network}
 \acro{WMAN}{wireless metropolitan area network}
 \acro{WNC}{wireless network coding}
 \acro{WRAN}{wireless regional area network}
 \acro{WSEE}{weighted sum of the energy efficiencies}
 \acro{WPEE}{weighted product of the energy efficiencies}
 \acro{WMEE}{weighted minimum of the energy efficiencies}
 \acro{X2}{x2}
 \acro{X2-AP}{x2 application protocol}
 \acro{ZF}{zero forcing}

 \end{acronym}

\end{document}